\documentclass[11pt]{article}

\usepackage{fullpage}
\usepackage{cite}
\usepackage{graphicx}
\usepackage{textcomp}
\usepackage{xcolor}
\usepackage{tabularx}
\usepackage{wrapfig}
\usepackage{listings}
\usepackage{multirow}
\usepackage{makecell}
\usepackage{caption}
\usepackage{pifont}  
\usepackage{amsmath}
\usepackage{subcaption}
\usepackage{enumerate}
\usepackage{threeparttable}
\usepackage{url}
\usepackage[hidelinks]{hyperref}
\usepackage[hyphenbreaks]{breakurl}

\usepackage{comment}

\begin{document}

\newcommand{\cmark}{\ding{51}}
\newcommand{\xmark}{\ding{55}}

\title{CRAC:  Checkpoint-Restart Architecture for CUDA with Streams and UVM}
\date{}

\author{Twinkle Jain\textsuperscript{*} and Gene Cooperman\thanks{\noindent
        This work was partially supported by National Science
        Foundation Grant OAC-1740218 and a grant from Intel Corporation.} \\
\textit{Khoury College of Computer Sciences} \\
\textit{Northeastern University}\\
Boston, USA \\
\{jain.t,g.cooperman\}@northeastern.edu
}

\maketitle

\maketitle

\begin{abstract}
The share of the top 500 supercomputers with NVIDIA GPUs is now over
25\% and continues to grow.  While fault tolerance is a critical issue
for supercomputing, there does not currently exist an efficient,
scalable solution for CUDA applications on NVIDIA GPUs.  CRAC
(Checkpoint-Restart Architecture for CUDA) is a
new checkpoint-restart solution for fault tolerance that supports the
full range of CUDA applications.  CRAC combines: low runtime overhead
(approximately 1\% or less); fast checkpoint-restart; support for scalable CUDA
streams (for efficient usage of all of the thousands of GPU cores); and
support for the full features of Unified Virtual Memory (eliminating
the programmer's burden of migrating memory between device and host).
CRAC achieves its flexible architecture by segregating application code
(checkpointed) and its external GPU communication via non-reentrant
CUDA libraries (not checkpointed) within a single process's memory.
This eliminates the high overhead of inter-process communication in
earlier approaches, and has fewer limitations.
\end{abstract}

\section{Introduction}
\label{sec:introduction}

General-purpose GPU computing continues to become more important in
supercomputers and in large- and medium-size clusters.  For example,
starting from zero GPUs in~2010, the
number of clusters with NVIDIA GPUs has reached 136 out of~500 in the
Nov., 2019 listing of the top 500~supercomputers~\cite{TOP500-2019},
as seen in the next graph.

\begin{center}
  \includegraphics[width=1.0\columnwidth]{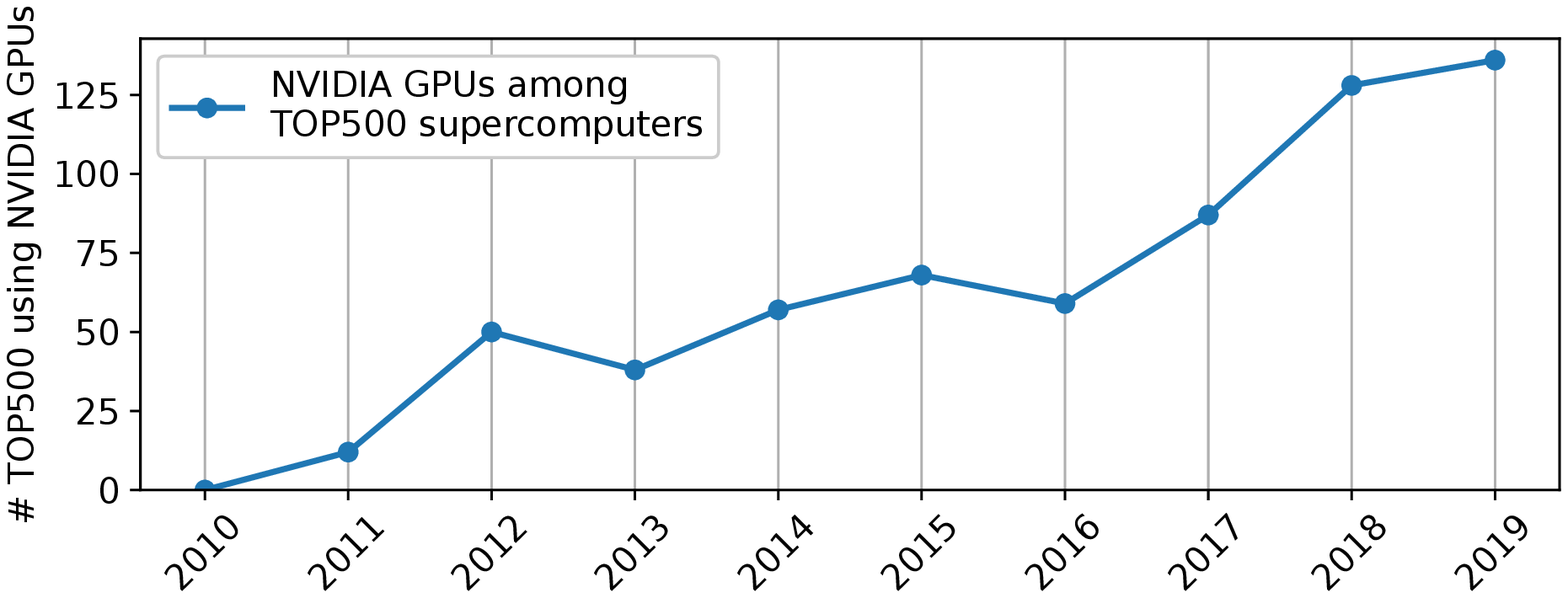}
\end{center}

This work introduces CRAC (Checkpoint-Restart Architecture for CUDA)
for transparently checkpointing CUDA on GPUs.  Transparent checkpointing
for CPUs (as opposed to GPUs) has long been important in long-running
computations.
Transparent checkpointing is widely available for Linux HPC applications,
including MPI.
Three notable examples of transparent checkpointing
are DMTCP~\cite{ansel2009dmtcp}
(multi-host and MPI), BLCR~\cite{hargrove2006blcr} (single-host
and MPI), and CRIU~\cite{criu} (primarily for single-host).
However, that ability to transparently checkpoint computations using GPUs
has been notably lacking.

Transparent checkpointing is important in HPC for at least four reasons:
\begin{enumerate}
\item[(a)] long-running batch jobs that might need more time to complete than
the typical 24-hour job allocation slot;
\item[(b)] fault tolerance (especially concerning GPU soft errors);
\item[(c)] backfill policies for efficient scheduling of batch queues; and
\item[(d)] process migration in the cloud,
for example to exploit spot instances in the cloud for cost-effective
computing~\cite{yi2011monetary}, and for other just-in-time strategies.
\end{enumerate}

The ability to checkpoint GPUs is even more pressing as clusters
and supercomputers continue to scale to an increased number of GPUs.
This is because of
the vulnerability of GPUs to soft errors.  A series of papers in
the literature has highlighted the issue of fault tolerance
for GPUs in the presence of soft errors~\cite{nie2016large,
sridharan2015memory,tiwari2015reliability,tiwari2015understanding}.
In particular, NVIDIA GPUs do not have the same level of error protection
of RAM as is the case for the high-end host computers used in clusters.

Finally, {\em transparent checkpointing} (as opposed to application-specific
checkpointing) is especially important in order to relieve the application
developer of the burden of coding for checkpointing.
There are several anecdotes in the community
of long-standing computational toolkits that ``used to'' have an
application-specific checkpointing module, but that specialized module
gradually became out-of-date as additional stateful parameters were
added to a model.

Further, application-specific checkpointing
typically has limitations, in that a checkpoint may be taken only
at each iteration of the outermost loop.  This is done in order to
avoid the complication of restoring the stack as it existed at runtime.
These limitations imply that application-specific checkpointing is often
incompatible with {\em on-demand checkpointing}, which is required
in the case of
spot instances, or when a large high-priority job arrives and existing
jobs must immediately be checkpointed.

Ironically, while the need for transparent checkpointing of GPUs
has grown in the last decade, the support for transparent checkpointing
of GPUs has diminished.
A series of results for transparent checkpointing of GPU~\cite{shi2009vcuda,gupta2009gvim,takizawa2009checuda,
gomez2010transparent,nukada2011nvcr} have stopped working as of CUDA~4.0.
This is because CUDA~4.0 introduced, in 2011,
 UVA (Unified Virtual Addressing between
host and GPU device).  This was later refined, with CUDA~6.0, to
UVM (Unified Virtual Memory).  All previous checkpointing efforts relied on
the ability to save and restore the CUDA library in memory.  But now
that the virtual memory address space is shared between GPU device
and host, any attempt to restore the checkpointed CUDA library
and associated allocated memory
at their original address will create inconsistencies between
the host and GPU device address space.

Two more recent efforts at checkpointing
(CRCUDA~\cite{suzuki2016transparent} and CRUM~\cite{garg2018crum}) try
to get around this problem by creating separate proxy processes.
CRCUDA presents a preliminary attempt whose overhead was apparently
never evaluated on real-world programs.
CRCUDA's github repo~\cite{crcudaSource} has not been active since~2015.
CRUM presents a more complete solution, but it continues to have limitations.

The problem with both CRCUDA and CRUM is that their approach
centers around
passing all CUDA calls from the application process to a CUDA library resident
in an independent proxy process.
This requires copying
buffers between the application process and proxy process before and after
each CUDA library call.  This has three inherent problems:
\begin{enumerate}
\item[(a)] Copying buffers creates a high runtime overhead.
        Modern CUDA applications may need to launch 1,000 CUDA kernels
        per second and more.
        (CRUM reports 6\%  to 12\%
        overhead~\cite[(Section~IV.B, figure~4(b))]{garg2018crum}.)
\item[(b)] CRUM's support for UVM is incomplete.
	The issue is that UVM allows for hardware-supported
	page faults between host and device whenever one or the other
	updates the memory in a unified page.
        CRUM is limited to supporting applications that follow this
        pattern:  CUDA-call, read from UVM, modify, write to UVM,
        next CUDA-call.  Not all applications follow this pattern.  See
	CUDA call~\cite[(Section~III.B)]{garg2018crum} for details.
\item[(c)] Neither CRCUDA nor CRUM appear to have been tested in
	checkpointing the maximum permitted number of concurrent
        CUDA streams.
        We speculate that the reason is that
	both approaches sustained a significant overhead in making
	a CUDA call, since this required copying memory buffers
        (arguments to the CUDA call) to an
	independent proxy process.  The essence of using CUDA Streams
	is to execute multiple CUDA kernels simultaneously (in multiple
	streams).  This parallelism implies a higher frequency of
	CUDA kernel calls, placing more stress on the memory transfers
	to the proxy process.
\end{enumerate}

In summary, this work makes three important contributions that may
be summarized as (a)~low runtime overhead, (b)~efficient support for UVM,
and (c)~efficient support for many concurrent CUDA streams.  More explicitly:
\begin{description}
  \item[\emph{1) Low runtime overhead:}]
    Previous checkpointing support for CUDA~4.0 and later had
    unacceptably high runtime overhead (for example, CRUM's 6\% to
    12\%~\cite{garg2018crum}).  The single-address space approach of
    this work enables more efficient, direct passing of pointers to
    CUDA kernels upon launch.  While doing this, it retains isolation of
    the CUDA application program from the helper (proxy) program that
    ``talks'' to the GPU.
  \item[\emph{2) Efficient and complete UVM support:}]
    There are no compromises in the UVM support.
    CRUM's shadow page synchronization restricts UVM-based
    applications solely to a single read-modify-write cycle between
    CUDA kernel launches~\cite[Section~III-B]{garg2018crum}.
    Further, CRUM's strategy fails when two concurrent CUDA streams
    write to the same
    memory page.
  \item[\emph{3) Many concurrent CUDA streams:}]
    The new approach scales well with many concurrent CUDA streams.
    The lack of previous experiments in the literature for more than
    two concurrent CUDA streams confirms the novelty of this work's
    support for many concurrent streams.
\end{description}

Finally, CRAC is free and open-source software.  The current version
of CRAC is found at:
\url{https://github.com/DMTCP-CRAC/CRAC-early-development.git}.
In the future, the newest version of CRAC will be included as a plugin
in the mainstream DMTCP~\cite{dmtcp_git}, which is open source.

In the remainder of this work, Section~\ref{sec:background} describes the
approach of three previous systems for transparent checkpointing of CUDA:
CheCUDA (basic approach), and CRCUDA, and CRUM (proxy-based approaches).
It also describes the deficiencies of those systems for use in HPC.
Section~\ref{sec:designAndImplementation} describes the new single
address-space approach of CRAC.
Section~\ref{sec:experimentalResults} presents experimental results
demonstrating the performance and generality of the new approach.
Section~\ref{sec:relatedWork} then describes the related work.
Section~\ref{sect:conclusion} presents the conclusion and future work.

\section{Background}
\label{sec:background}

We highlight the history of CUDA and earlier approaches to transparently
checkpoint CUDA applications.  This highlights why older approaches
stopped working with the introduction of CUDA-4.0, and a conceptually
new approach was required.

\subsection{The Historical Evolution of CUDA}
As described in the introduction, previous mechanisms for transparent
checkpointing~\cite{shi2009vcuda,gupta2009gvim,takizawa2009checuda,
gomez2010transparent,nukada2011nvcr} were made incompatible by the
introduction of Unified Virtual Addressing (UVA) in CUDA~4.0.
UVA was introduced in CUDA-4.0, and was later refined into
Unified Virtual Memory (UVM) in CUDA~6.0.
UVM operates in analogy with the introduction of virtual memory for UNIX.
The CUDA UVM-enabled hardware and software execute on-demand paging,
so that application programmers don't need to explicitly swap memory
segments in and out of the GPU device.  CUDA streams were introduced
with CUDA-3.0 (Fermi GPUs).

\subsection{A First Attempt at Checkpoint-restart:  CheCUDA
            prior to CUDA~4.0}
\label{sec:background-early-CUDA}

Here, we describe the architecture of
CheCUDA~\cite{takizawa2009checuda}, built upon CCUDA-2.2 in 2009, as
representative of the general approach.  The basic steps are:
(a)~to ``drain the queue'' of tasks (of pending CUDA kernels)
using \texttt{cudaDeviceSynchronize} or \texttt{cuCtxSynchronize};
(b)~to copy persistent GPU state associated with resources held by the CUDA
library to host memory; (c)~to destroy all CUDA resources; (d)~to checkpoint
on the host side using BLCR~\cite{hargrove2006blcr}; and to restart by
reversing these steps.  Creation of CUDA resources is recorded prior
to checkpoint time, and then restored during restart in a classic
log-and-replay strategy.

A problem was encountered with CheCUDA and related approaches for
checkpointing GPUs~\cite{shi2009vcuda,gupta2009gvim,takizawa2009checuda,
gomez2010transparent,nukada2011nvcr} in 2011.  This is the year when
NVIDIA introduced one more CUDA resource as part of the CUDA~4.0 library:
the unified virtual address (UVA) facility.  CUDA did not provide
an API to save the state of UVA and later restore it.  This was not
surprising, since the UVA resource is shared between device and host,
and so it would be difficult to provide a user API to restore it.
Previous CUDA resources were resident solely on the GPU.

\subsection{A Second Attempt at Checkpoint-restart:  Proxy-based
           solutions for CUDA~4.0 and later}
\label{sec:background-proxies}

In 2011,
CUDA~4.0 introduced UVA (Unified Virtual
Addressing)~\cite{schroeder2011peer}.  CUDA~6.0 then introduced UVM (Unified
Virtual Memory) in 2013~\cite{harris2013unified}, exacerbating further
the difficulty of saving and restoring UVA or UVM state.  UVM on Pascal
and later GPUs supports hardware page faulting of host pages into the
GPU and vice versa.

CUDA memory allocations were then a resource that could no longer be saved
and restored, since a memory allocation included a virtual memory mapping
between host and device.  That mapping is managed by the NVIDIA portion
of the operating system, and it was not exposed to the CUDA programmer.

To overcome this,
CRCUDA~\cite{suzuki2016transparent} and CRUM~\cite{garg2018crum}
took a proxy-based approach.  But CRCUDA doesn't
support UVA or UVM.  CRUM supports UVM through \emph{shadow
memory}~\cite[Algorithm~1]{garg2018crum}, but at the cost of high runtime
performance, and covering only standard CUDA applications following
the read-modify-cudaCall pattern.

\section{The Design and Implementation of CRAC}
\label{sec:designAndImplementation}

CRAC provides the ability to save and restore the state of
CUDA by first using CUDA-specific save/restore operations, and then
delegating to a traditional checkpoint-restart package.  Conceptually,
CRAC could have used any of the three most popular
systems for transparent checkpointing: BLCR~\cite{hargrove2006blcr},
CRIU~\cite{criu}, and DMTCP~\cite{ansel2009dmtcp}.  However,
CRIU does not support checkpointing of multiple hosts and BLCR
is no longer actively maintained.  In the end,
the support of DMTCP for process virtualization and
plugins~\cite{arya2016design} makes it easier to add modular
support for CUDA without having to excessively understand details of
the internals of the host checkpointing package.
Further, DMTCP remains the only transparent checkpointing
package to operate at petascale, as originally demonstrated
in~2016~\cite{cao2016system}, when it was used to checkpoint two petascale
computations:  MPI-based HPCG (using 32,752 cores) and MPI-based NAMD
(using 16,368 cores)~\cite{cao2016system}.

The discussion of CRAC is next split into two parts: design and
implementation.

\subsection{The Design of CRAC}
\label{sec:design}

The problems with previous approaches to transparently checkpointing
CUDA using proxies were highlighted in the introduction and
Section~\ref{sec:background}:  high runtime overhead due to inter-process
communication; and the difficulty of supporting certain newer CUDA resources
(e.g., UVA/UVM and multiple CUDA streams) when the CUDA API did not
expose a mechanism for easily saving and restoring those resources.

The inter-process communication bottleneck between a
CUDA application process and a proxy process is an essential bottleneck
of CRCUDA and CRUM.
CRAC's solution is to combine the application and proxy into a
single process, whose address space contains \emph{two independent
programs, each with their own text segment, data, heap, and
runtime libraries}.  The application program and a proxy program
are loaded separately into the same address space, where the Linux kernel
views them as a single process.
Yet, the application and the proxy (now renamed
a helper program) are linked to two, independent runtime libc libraries
and two runtime loaders (ld.so).

\begin{figure*}[ht]
  \centering
  \includegraphics[width=0.6\textwidth]{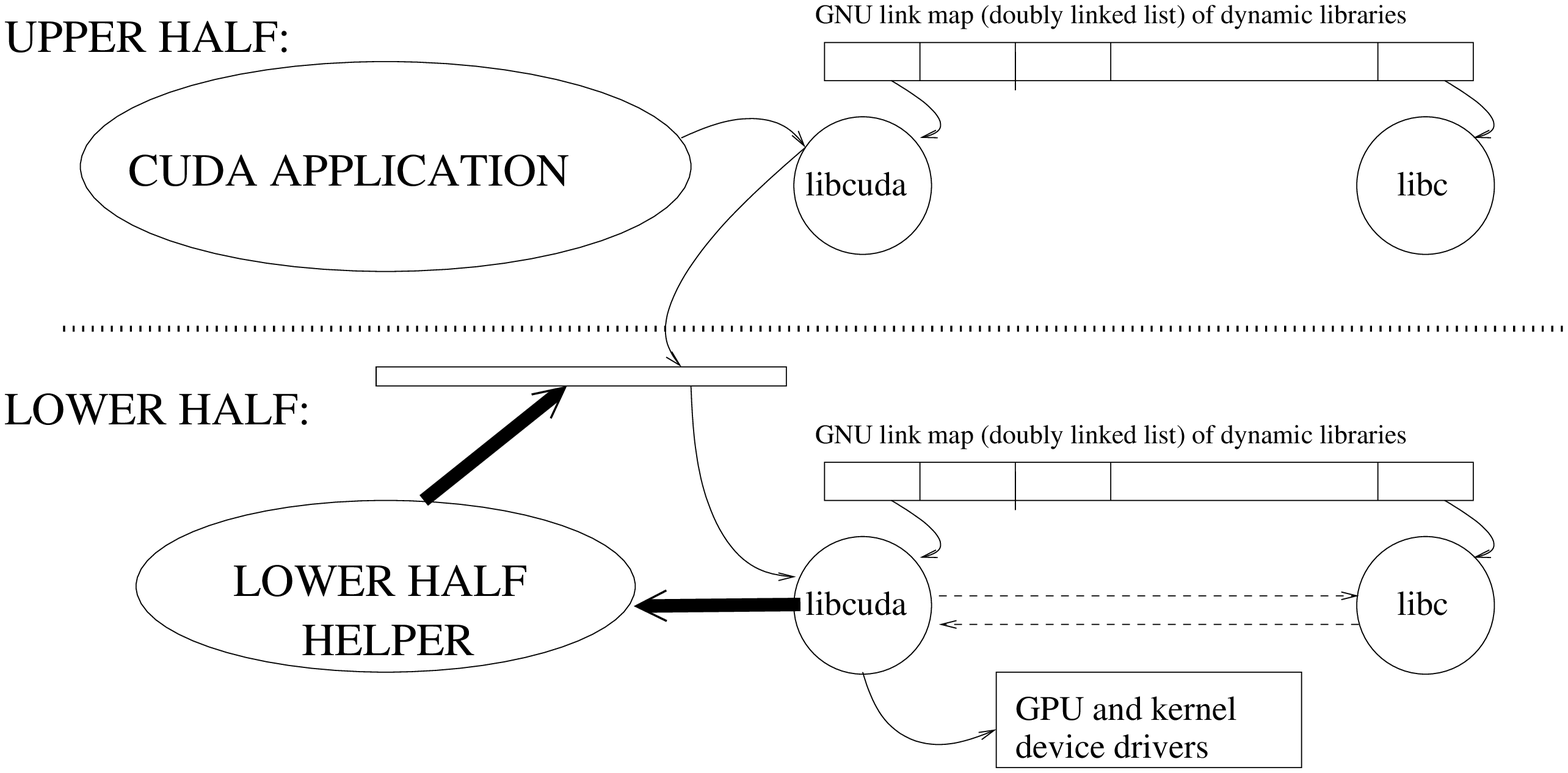}
  \caption{ \label{fig:split-process}} \small\raggedright
  \textbf{Split Processes:}
  The lower-half helper program is a tiny CUDA application that was
  loaded into the ``lower half'' of the virtual memory address space.
  At the time of launch, it copied the entry points of CUDA library
  calls from the lower-half libcuda to an array of libcuda entry addresses.
  When the main CUDA application was launched (in the upper half),
  it was launched under control of DMTCP.  DMTCP arranged to create
  a trampoline from the upper-half libcuda to the lower-half
  libcuda entry point, via the libcuda addresses found in the array
  created by the lower-half helper program.  Now, at runtime, when
  the end user's CUDA application makes a call to the CUDA library,
  the trampoline causes control to be passed to the lower-half
  libcuda.  Later, at checkpoint time, only the memory in the
  upper half will be saved.  At restart time, a new lower-half
  CUDA program is loaded into memory, and it re-initializes the
  array of libcuda addresses.  It then restores the upper-half
  memory from the checkpoint image, and passes control back
  to the CUDA application. Note that libcuda is a representative term for CUDA runtime
  library here.
\end{figure*}

The application program that was loaded into memory
is linked to a dummy CUDA library
that passes all CUDA calls
to the proxy program that was loaded.
The proxy program contains the active CUDA library,
and only the code of the proxy program that communicates with the GPU.
For a diagram illustrating the relationship,
see Figure~\ref{fig:split-process}.

Checkpoint and restart then proceed more or less as described in
Section~\ref{sec:background-early-CUDA} (CheCUDA prior to CUDA~4.0).
However, there is a crucial distinction.  We do not save the memory
of the proxy program.  Hence, we are not saving the memory of the
active CUDA library that talks to the GPU.  The CUDA library includes
stateful memory associated with CUDA resources such as UVA/UVM-based
memory.

On restart, we will load a completely new copy
of the proxy program.  The CUDA library of the new copy of the proxy
has its original state.  The stateful memory of the CUDA library
is put back in its initial state.  This new architecture again
makes feasible the classical log-and-replay of CheCUDA and other
applications.  The use of log-and-replay in CRAC is described fully
later in this section.

The literature describes two ways to implement this single address-space
design:
\emph{split processes}~\cite{garg2019mana} (loading two programs in the
same address space) and \emph{process-in-process}~\cite{hori2018process}
(using Linux's \textit{dlmopen} to offer independent namespaces,
Using dlmopen is superficially attractive, due to the greater
simplicity of this approach.  Therefore, we analyze this case first.

\setcounter{paragraph}{0}
\paragraph{Single address-space design:  process-in-process}

Process-in-process~\cite{hori2018process} was introduced as a mechanism
to reduce the overhead in inter-process communication between two
MPI ranks (processes) that coexist on the same host.  By placing the two ranks
within a single process by using \texttt{dlmopen}, runtime overhead was reduced.
It became possible to directly pass pointers between the two MPI ranks, instead
of relying on inter-process communication techniques.

This simple approach is attractive, and it captures many of the goals
of split processes, as depicted in Figure~\ref{fig:split-process}.
However, this approach is not conducive to our requirement of tracking
memory associated with the CUDA application program, versus the
helper program.  The NVIDIA compiler ({\tt nvcc}) links both the CUDA
application and the helper program with several libraries --- in particular,
the NVIDIA CUDA library and the runtime library.  It becomes difficult
to associate each memory region according to whether it was loaded by
a library for the CUDA application or a library for the helper program.

\paragraph{Single address-space design:  split processes}
\label{sec:split-processes}

Split processes~\cite{garg2019mana} were introduced as a mechanism
to separate the MPI and network libraries from the end user's
MPI application.  The split-process approach is more or less the same
as described for CRAC near the beginning of Section~\ref{sec:design}.
There is an important distinction in that in the case of MPI, the proxy
or helper program was statically linked~\cite[Section~3.6]{garg2019mana}.
NVIDIA encourages CUDA programs to be linked dynamically.  Even when
the \texttt{-static} flag is passed to NVIDIA's compiler, some NVIDIA
libraries remain dynamically linked.

In this scheme, the helper program is loaded
first, resulting in a new process.  That process then directly loads
the CUDA application into memory.
It is important to track all memory
allocations (all calls to mmap) by the lower half, so that they
are not checkpointed.  To accomplish this,
a program loading mechanism is used that imitates
the way in which the kernel loads an application.  (The kernel
first loads an ELF interpreter into memory, since the ELF interpreter
is structured as a statically linked executable with text, data, and stack.
The ELF interpreter then loads the dynamically linked target executable.)
The loading mechanism is modified to interpose on all calls to {\tt mmap()}.
This allows our kernel loader to load each memory region
(including the several NVIDIA libraries) into
a restricted portion of the address space, using the {\tt MAP\_FIXED}
parameter.  This approach also yields the illustration in
Figure~\ref{fig:split-process}, but it provides a mechanism for
associating each memory region conceptually with an ``upper-half''
or ``lower-half'' portion of the address space.

\paragraph{Log-and-replay}

Prior to CUDA-4.0, copying the persistent state of CUDA was exemplified
by copying two allocation arenas:
\texttt{cudaMallocHost} on the host;
and
\texttt{cudaMalloc} on the device, or GPU.
Just prior to a checkpoint, the data on host and device was copied
to a special location, and it was restored on restart.
However, CUDA-4.0 and later introduced \texttt{cudaMallocManaged} for
managed memory, used with UVM.  CRCUDA cannot support UVM
at all, and CRUM supports it imperfectly, as described in
Section~\ref{sec:background-proxies}.

Hence, copying the full persistent state at checkpoint time has become more
of a challenge since CUDA-4.0.  CheCuda and earlier approaches had destroyed
any CUDA resources prior to checkpointing, and restored them on resume and
restart.
This worked because the persistent resources of the CUDA library prior to
CUDA~4.0
could be logged and later restored.
With the advent of UVA/UVM in CUDA-4.0 and later, the unified virtual
memory is an essential resource that could not be recovered once
destroyed.  It appears from our own experiments that the UVM resource
had permanently modified the memory of the CUDA library's state, and the
restored
CUDA library was then inconsistent when called after restart.

Copying the persistent state would require reverse-engineering the
CUDA library, which is all but impossible, due to
the closed-source nature of CUDA.  But the CUDA library
has internal bookkeeping information on the contents of those three
allocation arenas.  Upon restart, each allocation must be recreated
at the original lower-half address that existed prior to checkpoint.

By interposing on the cudaMalloc family of CUDA calls, a log-and-replay
approach is used by CRAC to
copy to the upper half and later restore the memory regions,
in the same order as when they were allocated.
This takes advantage of
CUDA's internal deterministic bookkeeping for the allocation arenas.
On restart, a fresh CUDA library in the lower half would allocate
the memory regions at the same addresses as originally seen.

This traditional log-and-replay approach described in
Section~\ref{sec:background} is compatible with split processes only when
targeting smaller CUDA applications.  But this widely used log-and-replay
approach is observed to fail on more complex applications.  It fails
for two reasons.  In order to apply the approach faithfully and take
advantage of determinism in the CUDA library, it would be necessary to
re-execute (replay) in the original ordering all calls in the family
of cudaMalloc and cudaFree.  Second, this approach becomes more difficult
when supporting concurrent streams, since two threads on the host may
concurrently make calls to cudaMalloc, which would require an extra
global lock on all calls to cudaMalloc in the lower-half library.
The next section discusses the memory management approach
actually used by CRAC.

\subsection{Implementation Issues}
\label{sec:implementation}

Having chosen the split process approach for CRAC, there were
several implementation issues arising for the case of CUDA that
were not present in the case of MPI.

\subsubsection{Implementation:  Issue of library-allocated memory}

The largest complexities of adapting split processes from MPI to CUDA
arise from the differing conventions of allocating memory.
The design of MPI assumes that calls to MPI will employ
caller-allocated memory: callers to the MPI library pre-allocate buffers
and pass them to MPI.

The design of CUDA assumes \emph{callee-allocated}, or library-allocated
memory:  the CUDA library in the lower half may allocate its own
internal buffers, and then
return those buffers to the calls.  A good example is \texttt{cudaMalloc}
to allocate host memory for the application.
This CUDA routine allocates its own memory, and potentially invokes
\texttt{mmap} to do so.

One can argue that an mmap call can be intercepted, in order to do
deterministic replay.  However, we observed that a single cudaMalloc
call can make many calls to mmap.  Moreover, the first cudaMalloc will
create a large CUDA malloc arena through mmap.
This mmap call may fall into the middle of several other mmap calls.
Subsequent cudaMalloc call might not call mmap at all.
This results in two problems.  (a)~It is impractical to interpose on many
mmap calls in order to identify the particular mmap calls of interest to us.
(b)~The active CUDA malloc buffers to be checkpointed will generally be
a small fraction of the full CUDA malloc arena that was created.

To counter this problem, we log only the host or device pointers to
buffers that were created by a call from the cudaMalloc family of APIs.
This helps CRAC in improving performance by avoiding unnecessary interceptions
in the lower-half.

\subsubsection{Implementation: Issue of memory overlapping}

The lower- and upper-half memory regions can appear anywhere in the
process address space. In DMTCP, one part of saving the state of a running
process includes reading the /proc/PID/maps and saving memory regions. In
/proc/PID/maps, two memory regions with the same permissions get merged
after allocation.  This makes it harder to decide if whole or part of a
memory region belongs to the upper half and must be checkpointed. This
has not been an issue in the case of MANA for MPI~\cite{garg2019mana},
where the lower half is compact since it is compiled as a statically
linked executable.

Another issue that we observed is that when the library of the lower
half allocates memory pages, it may overwrite the upper-half's existing
memory pages, and indeed, it may even unmap some of the upper-half's
existing memory pages. This could lead to silent memory corruption.

To counter these problems, CRAC tracks all the allocations done by the
upper half and also tries to consolidate memory regions created by
the upper half, as described in~\ref{sec:split-processes}.

\subsubsection{Implementation: Saving the ``library-allocated'' arena}

Since CRAC can interpose on the CUDA library in the lower half,
it can interpose on all calls to {\tt mmap()}.  Naively,
one would assume that for each of the cudaMalloc family of calls,
there is a single call to {\tt mmap()}, which can be recorded and
replayed. This does not work, since a cudaMalloc call may make
multiple calls to {\tt mmap()}. Or a single cudaMalloc call may
use {\tt mmap()} to create a large memory region that acts as
an allocation arena for later calls to cudaMalloc. While this is helpful for the
CUDA library's memory management algorithm, it is not desirable to save the
entire arena --- especially, when cudaMallocs actually uses only a small
portion of the allocation arena.

To counter this, CRAC does its own internal bookkeeping.
Rather than saving a large allocation arena that makes the checkpoint size
larger unnecessarily, we only save the memory associated with active mallocs.
Active mallocs are those allocations that were allocated but not freed at the
time of checkpoint. Draining and refilling device memory at active mallocs is
essential to make the device state consistent across checkpoint and restart.
Note that saving the memory associated with the active mallocs is different
than logging the sequence of all cudaMallocs and all cudaFrees. While we save
only the memory associated with active mallocs, we still need to replay the
entire original sequence to get the same host and device addresses as prior
to checkpoint (explained in the next section).

\subsubsection{Implementation: restoring the CUDA library-allocated regions}

An important implementation issue for CRAC is to restore all of
CUDA's memory allocations at their original address during restart.
CUDA has three primary allocation routines: cudaMalloc (on the device),
cudaMallocHost/cudaHostAlloc (on the host), and cudaMallocManaged (for UVM: unified memory
on device and host).  CRAC logs all CUDA calls that allocate and free memory.

In the case of cudaHostAlloc, it suffices to keep track of only the
{\em active} memory buffers (the buffers that have not been freed at the time
of checkpoint).  At restart time, CRAC only needs to replay cudaHostMalloc
for active memory buffers, in order to again register these buffers with
the CUDA library.  Note that the memory buffers are already present in the
restored upper half memory.

In the cases of cudaMallocHost (on the host), cudaMalloc (on the device)
and cudaManagedMalloc (for unified memory), CRAC replays all associated
allocations and frees at restart time.  The memory associated with these
regions is saved at checkpoint time and copied back at restart time.
In our experiments on real-world applications, we observed many calls
to cudaMalloc and cudaManagedMalloc, but few calls to free those buffers.

CRAC replays the entire log in order to guarantee that active memory
allocations are restored at the original address.  CRAC relies on determinism
of the CUDA library allocation.  CRAC also disables address space randomization
using Linux's \texttt{personality} system call.
And CRAC's determinism also relies on using the same CUDA/GPU
platform on restart.  In the future, three possible solutions can be
implemented to optimize this:  virtualization of library-allocated
addresses; patching applications locations containing the addresses;
or a future enhancement by NVIDIA offering a MAP\_FIXED flag analogous
to the flag of the mmap call.

\subsubsection{Implementation: Handling CUDA's internal registration of
fat binaries:}

At the time of launching a CUDA application, CRAC must
arrange for the CUDA library in the lower half to register the CUDA
kernels residing in the upper half as the active CUDA library loads before the
upper half.  This requires that CRAC call the lower level CUDA functions in the
lower-half CUDA library:
\texttt{\_\_cudaRegisterFatBinary}, \texttt{\_\_cudaRegister<CUDA-element>},
and
\texttt{\_\_cudaUnregisterFatBinary} (during
cleanup at process exit). Here, CUDA elements are device variable, functions,
texture, surface, and etc. Finally, during restart, CRAC must
re-register the application kernels, since this is a fresh copy of the
lower half.  This may require additional patching of fat-binary-handle at
restart time. This added burden never occurs in the case of MANA for MPI:
As before, MANA for MPI benefits from the MPI standard, which defines an
almost complete isolation of the MPI library from the MPI application.

\section{Experimental Results}
\label{sec:experimentalResults}

This section present the comprehensive
analysis of the CRAC's performance for real-world applications.  The aim of
this section is to demonstrate that CRAC has low runtime overhead
and scales well on real-world applications.

\subsection{Hardware}
The experiments presented in this section are
performed on the \textit{PSG} cluster of NVIDIA. Each node runs
CentOS~7.7 release (kernel version 3.10.0), with 4~NVIDIA Tesla~V100 (compute
capability~7.0),
each with 32~GB of RAM. Each Haswell node is running two 16-core Intel
Xeon E5-2698~v3 (2.30~GHz) processors with a total of 256~GB of RAM.

A local, NVIDIA Quadro K600 node with 1~GB of RAM was used only
in Section~\ref{sec:fsgsbase}.  This section includes a special
set of experiments to analyze any runtime improvement using
the FSGSBASE patch to the Linux kernel~\cite{fsgsbase-v9}.
The FSGSBASE patch is under active review for inclusion in the
mainline Linux kernel~\cite{fsgsbase-lwn,fsgsbase-v13}.  A custom
Linux kernel version~5.0.6 was built from the official Ubuntu git
repository\footnote{git://kernel.ubuntu.com/ubuntu/ubuntu-disco.git}
on Ubuntu~18.04.3~LTS (Bionic Beaver).

\subsection{Software}
Each GPU runs NVIDIA CUDA version~10.0 with driver
440.33.01. We use NVCC to compile the application and use gcc/g++
version~7.3.0 for the linking with libc version~2.17. We use MPICH
version~3.3.2 for the MPI-based applications.

The experiments use CRAC, a DMTCP plugin~\cite{arya2016design}, developed
to specifically checkpoint and restart CUDA applications using the novel
split-process and user-space program loading mechanism
(Section~\ref{sec:split-processes}).
DMTCP~\cite{ansel2009dmtcp} is an open-source
tool for transparent checkpoint-restart for distributed and multi-threaded
applications. We use DMTCP version~3.0\cite{dmtcp_git} (master branch).

\subsection{Terminology} We define the following terminology and formulas that
will be used for the rest of the paper.
\begin{enumerate}[(a)]
\item Runtime
      overhead: we use the standard formula to calculate the runtime overhead
      where $E_{CRAC}$ is the execution time of an application under
      CRAC and
      $E_{\overline{CRAC}}$ is the native execution time (not under CRAC).
 \begin{equation}
  \textrm{Runtime Overhead \%} = \frac{E_{CRAC} -
  E_{\overline{CRAC}}}{E_{\overline{CRAC}}} \times 100
 \end{equation}

\item CUDA calls-per-second (CPS): CUDA API calls are calculated by NVIDIA's
  profiler
    \texttt{nvprof}. We are interested
    only in the number of calls from upper half to lower half,
    for the sake of analyzing their overhead.
    A simple script extracted just those calls from the upper half
    (i.e., to the lower-half CUDA runtime library), and not
    the calls to the CUDA device library (made directly from the lower-half
    CUDA runtime library).  There are three additional
    CUDA calls that the upper half can make:
    \texttt{cudaLaunchKernel} (reported by \texttt{nvprof}),
    along with two undocumented internal APIs,
    \texttt{\_\_cudaPushCallConfiguration}
    and \texttt{\_\_cudaPopCallConfiguration}.
    The CUDA compiler generates all three calls for one CUDA kernel launch.
    So, the formula for total CUDA calls is as follows:
    \begin{align*}
      \text{Total CUDA calls} =\,\,&3 \times count(\textrm{cudaLaunchKernel})
                                                                           + \\
                                   &count(\textrm{rest of CUDA runtime API})
    \end{align*}

  The CUDA ``calls per second'' (CPS) is defined as:
  \begin{equation}
    CPS = \frac{\text{Total CUDA calls}}{E_{\overline{CRAC}}}
                                                 {\label{eq:CUDA_calls}}
  \end{equation}
\end{enumerate}

\subsection{Application benchmarks}

CRAC is analyzed using six CUDA applications. Four
of them are standard benchmark suites or real-world applications,
and the rest are taken from the official NVIDIA CUDA reference code
suite\footnote{https://docs.nvidia.com/cuda/cuda-samples/index.html}.
These applications are chosen to cover a wide range of CUDA features,
including Unified Virtual Memory (UVM) and CUDA Streams.

\begin{table}[ht!]
  \centering
  \begin{tabular}{ |l|c|c|r|c| }
    \hline
    \textbf{Application} & \textbf{UVM} & \textbf{Streams} & \textbf{CPS}
    & \textbf{\# streams} \\ \hline
    \texttt{Rodinia}       & \xmark & \xmark & 38--132K & --- \\ \hline
    \texttt{Lulesh}        & \xmark & \cmark & 2.5K & 2--32    \\ \hline
    \texttt{simpleStreams} & \xmark & \cmark & 10K  & 4--128 \\ \hline
    \texttt{UnifiedMemory} & \cmark & \cmark & 4.4K & 4--128 \\
    \texttt{\hfill Streams}&        &        &      &        \\ \hline
    \texttt{HPGMG-FV}      & \cmark & \xmark & 35K  & ---    \\ \hline
    \texttt{HYPRE}         & \cmark & \cmark & 600  & 1--10  \\ \hline
  \end{tabular}
  \caption{Application benchmarks characterization}
  \label{tab:applications-characterization}
\end{table}

Table~\ref{tab:applications-characterization} characterizes the applications
used here. The table includes four
columns: UVM and Streams are checked if the application uses the respective
CUDA feature. The CUDA calls per second (CPS) are calculated using
equation~\ref{eq:CUDA_calls}.
Lastly, if the application uses CUDA streams, then the table indicates the
range of how many CUDA streams can be used with the application.

The Rodinia benchmark suite~\cite{che2009rodinia} provides a wide range
of applications with a varying CPS.
Also, two stream-oriented codes from the NVIDIA CUDA toolkit are used:
simpleStreams and UnifiedMemoryStreams~\cite{nvidia_code_sample}.
The two applications are chosen because: they exclusively demonstrate
the streams feature; and they can be configured easily to use the
maximum number of streams on a given GPU (128~streams in our case).

To evaluate CRAC's performance on real-world applications, we use three
benchmarks from the DOE
(Department Of Energy): Livermore Unstructured Lagrangian Explicit
Shock Hydrodynamics (LULESH) version~2.0~\cite{LULESH2:changes};
HYPRE: Scalable Linear Solvers and Multigrid Methods library
version~2.13.0~\cite{hypre}; and
High-Performance Geometric MultiGrid HPGMG (using the
Github repository's master
branch~\cite{hpgmg}).

\subsubsection{\textbf{Rodinia Benchmark Suite}}
Rodinia\cite{che_rodinia_2009,rodiniaweb} is a commonly used benchmark
suite for CUDA.  Version~3.1 covers
a diverse range of 23~CUDA applications using basic CUDA features,
and compatible with all CUDA versions starting from CUDA~version~2.x.

We use 14 of the applications from the Rodinia benchmark suite for this
work. The other 9~applications were omitted either because they were
too short (completing within one second), or because they are similar
to benchmarks already included in terms of the total number of
CUDA API calls, or because the total number of CUDA API calls was too few.

Rodinia's applications can be scaled by adjusting the input. We use
the command line arguments given in Table~\ref{tab:command-line-args}
for the respective Rodinia benchmark applications.

\begin{table}[ht]
  \centering
  \begin{tabular}{ |l|l| }
    \hline
    \textbf{Application} & \textbf{Command-line argument(s)}\\ \hline
    \texttt{BFS        } & graph1MW\_6.txt \\ \hline
    \texttt{CFD        } & fvcorr.domn.193K \\ \hline
    \texttt{DWT2D      } & rgb.bmp -d 1024x1024 -f -5 -l 100000 \\ \hline
    \texttt{Gaussian   } & -s 8192 -q \\ \hline
    \texttt{Heartwall  } & test.avi 104 \\ \hline
    \texttt{Hotspot    } & temp\_512 power\_512 output.out \\ \hline
    \texttt{Hotspot3D  } & 512 8 1000 power\_512x8 temp\_512x8 output.out \\
    \hline
    \texttt{Kmeans     } & kdd\_cup -l 1000 \\ \hline
    \texttt{LUD        } & -s 2048 -v \\ \hline
    \texttt{Leukocyte  } & testfile.avi 500 \\ \hline
    \texttt{Myocyte    } & 500 1 0 \\ \hline
    \texttt{NW         } & 40960 10 \\ \hline
    \texttt{Particlefinder} & -x 128 -y 128 -z 10 -np 100000 \\ \hline
    \texttt{SRAD       } & 2048 2048 0 127 0 127 0.5 1000 \\ \hline
    \texttt{Streamcluster} & 10 20 256 65536 65536 1000 none output.txt 1 \\
    \hline
    \texttt{LULESH} & -s 150 \\ \hline
  \end{tabular}
  \caption{Command-line arguments for Rodinia Benchmarks}
  \label{tab:command-line-args}
\end{table}

\begin{figure}[ht!]
  \centering
  \begin{minipage}{0.45\textwidth}
    \centering
    \includegraphics[width=\textwidth]{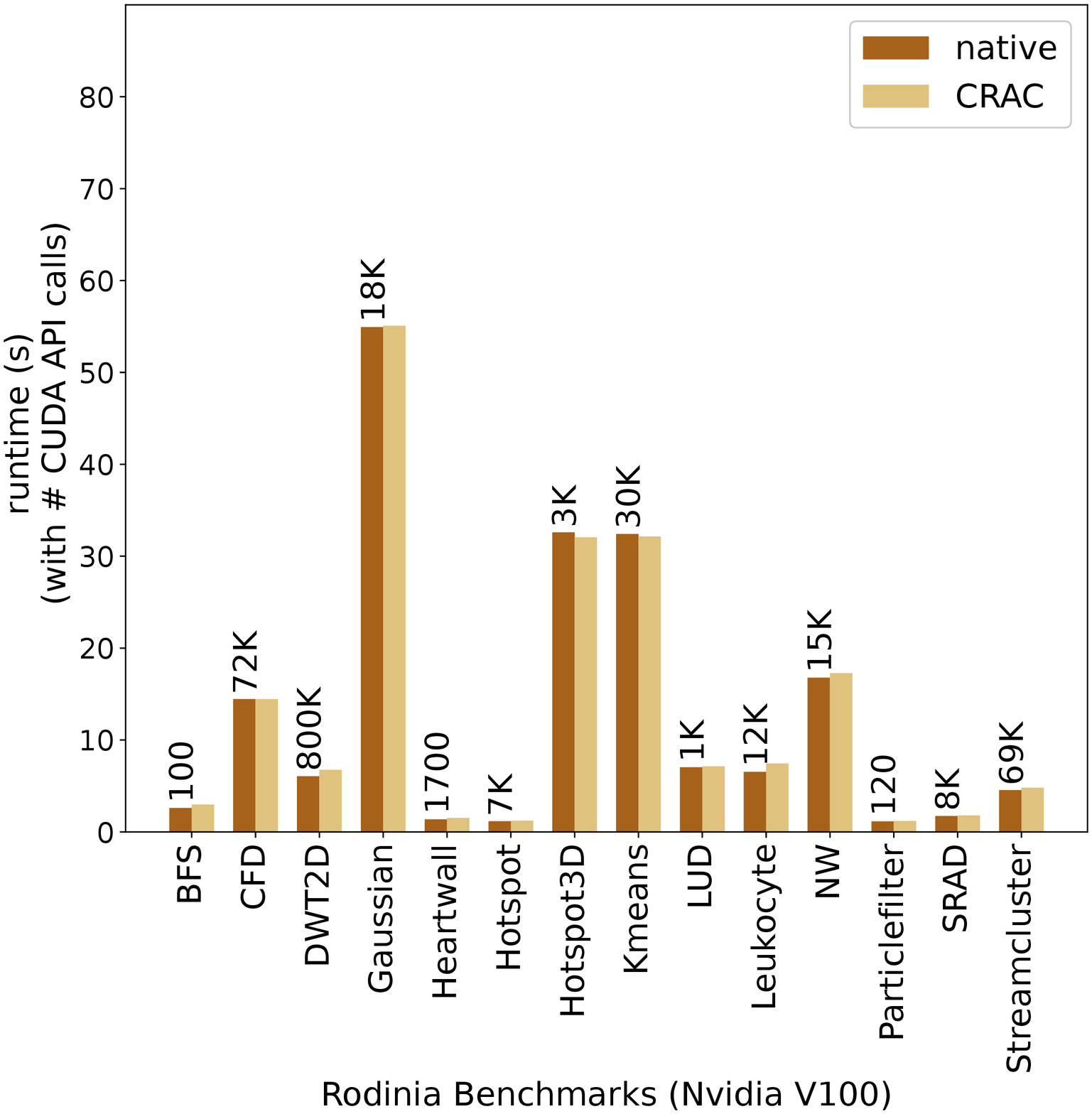}
  \caption{\label{fig:rodinia_runtime}
           Runtimes of 14~Rodinia Benchmarks with total CUDA API
           calls made by each benchmark (rounded off)}
  \end{minipage}
  \hspace{0.03\textwidth}
  \begin{minipage}{0.45\textwidth}
    \centering
    \includegraphics[width=\textwidth]{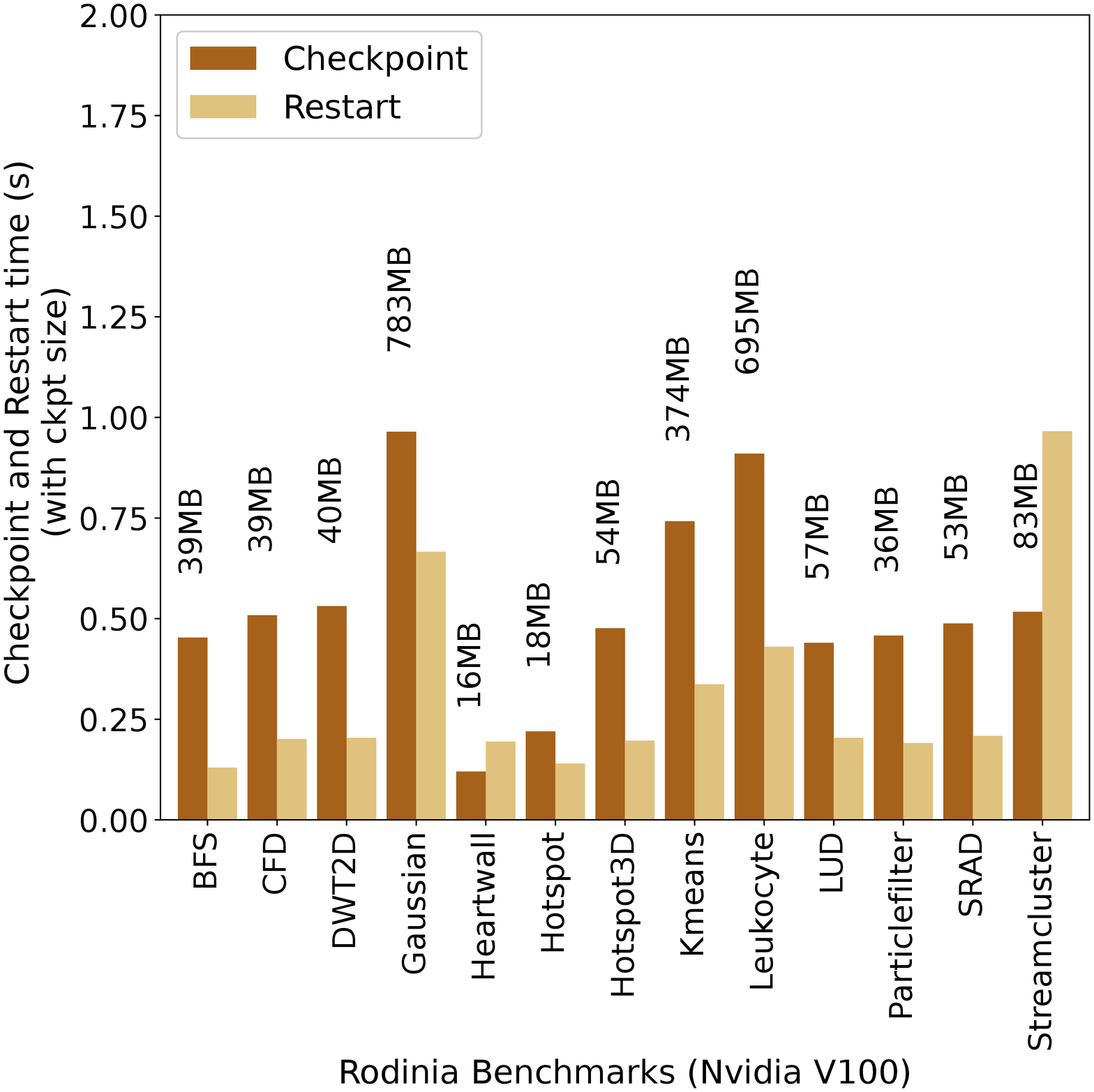}
  \caption{\label{fig:rodinia-ckpt}
           Checkpoint and restart times of 14~Rodinia benchmarks with
           checkpoint image sizes}
  \end{minipage}
\end{figure}

\begin{figure*}[ht]
  \centering
  \begin{subfigure}{0.40\textwidth}
    \centering
    \includegraphics[width=\textwidth]{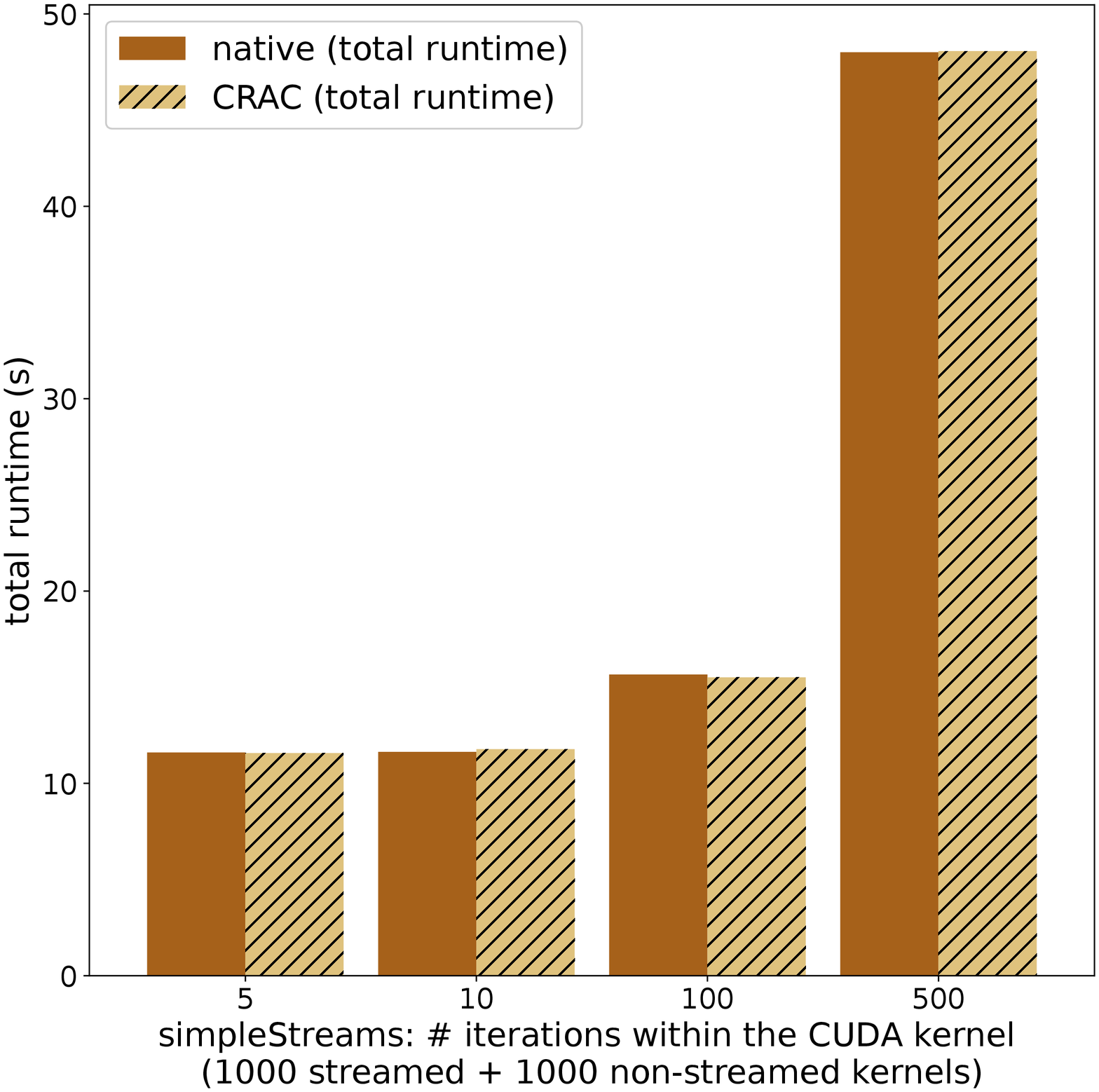}
    \caption{\label{fig:simpleStreamRuntimes}
             Runtimes of simpleStreams without and with CRAC while
             increasing the iterations within the CUDA kernel}
  \end{subfigure}
  \hspace{0.5in}
  \begin{subfigure}{0.40\textwidth}
    \centering
    \includegraphics[width=\textwidth]{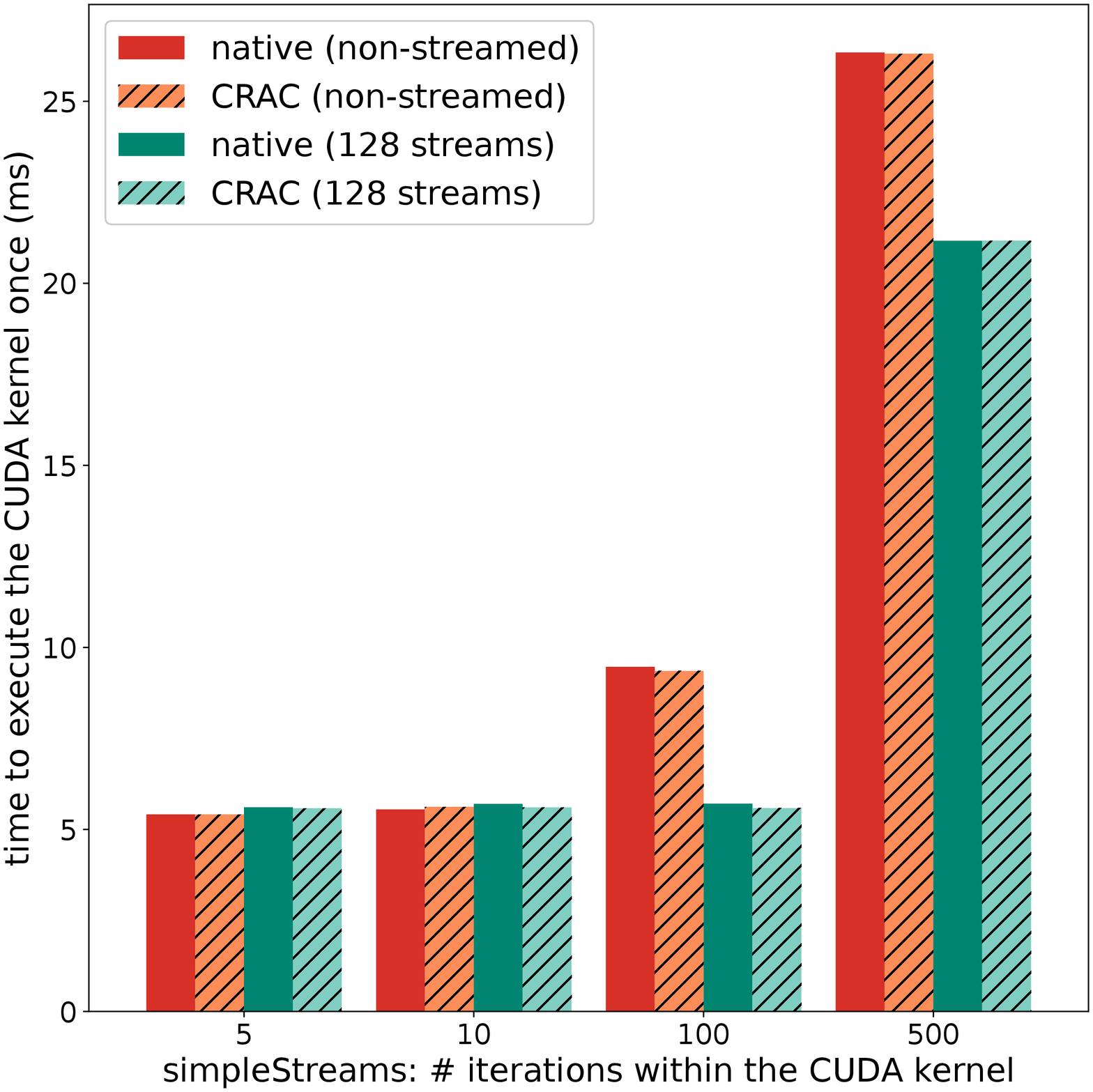}
    \caption{\label{fig:simpleStreamCUDAcalls}
             Time for one CUDA kernel execution time without and with
             streams in simpleStreams.  (More iterations imply
             a longer-running kernel.)}
  \end{subfigure}
  \caption{\label{fig:simplestream2figs}
           Experiments with simpleStreams from the NVIDIA CUDA code sample}
\end{figure*}

\paragraph{Runtime overhead}
    Figure~\ref{fig:rodinia_runtime} shows
    the runtimes of Rodinia benchmarks without CRAC (native) and with CRAC. We
    ran 10~iterations of each benchmark and calculated the mean for the
    each runtime. In almost every case,
    the 10~iterations had a standard deviation of approximately 0.1~seconds.
   \par
    \hbox{~~~~}The figure also shows
    that 9 out of 14~benchmarks namely, BFS, DWT2D, Heartwall, Hotspot,
    LUD, Leukocyte, Particlefilter, SRAD, and Streamcluster, ran in less than
    7~seconds. With these benchmarks, the runtime overhead varied between
    1\% and 14\%. There are two reasons for a higher overhead: first, with
    short-running applications, DMTCP's startup time becomes significant;
    second, with short-running tasks, the small standard deviation
    of 0.1~seconds becomes significant compared to the running time, and
    statistically leads to a higher overhead.
   \par
    \hbox{~~~~}On the other hand, the remaining Rodinia benchmarks run
    for more than 10~seconds, and we observe there a 0--2\% overhead.
    Interestingly, Hotspot3D,
    and Kmeans even have a negative overhead. We suspect that this is a result
    of caching.  Finally, CFD and Gaussian have less
    than 1\% overhead, while LUD and NW have less than 2\% overhead.

\paragraph{Checkpoint overhead}
    For checkpoint and restart, we disabled
    DMTCP's default \texttt{gzip} compression and triggered checkpoint
    at random times during an entire run of an application.
    Figure~\ref{fig:rodinia-ckpt} shows that
    the checkpoint-restart time is fairly
    small for CRAC and completes within one second for almost all
    cases. Checkpoint time is usually smaller than restart time,
    but there are two outliers (Streamcluster and Heartwall)
    for which restart takes more time than the checkpoint
    time. Further investigation showed that these two benchmarks specifically
    do many CUDA mallocs and CUDA frees. We log CUDA mallocs and frees to
    make the CUDA library's state consistent, and later replay those APIs on
    restart. We log the API when a user application calls the CUDA APIs that
    need to be logged. So, at checkpoint time, no extra work is
    needed, but at restart, those allocation and free calls were replayed.
    Note that even then the restart time is still less than 1~second.

\subsubsection{\textbf{Stream-oriented Benchmarks}}

\paragraph{simpleStreams}
SimpleStreams is one of the two code samples we took from NVIDIA's
official CUDA code samples. We quote from the code's documentation
that simpleStream illustrates the usage of CUDA streams for overlapping
kernel execution with device/host memcpy (memory copy).  The kernel is used
to initialize an array to a specific value, after which the array is
copied to the host (CPU) memory.  To increase performance, multiple
kernel/memcopy pairs are launched asynchronously, with each pair in its
own stream. Kernels are serialized.  Thus, if $n$~pairs are launched, a
streamed approach can reduce the memcopy cost to $(1/n)$th of a single copy of
the entire data set.

\begin{figure*}[ht]
  \centering
  \begin{subfigure}{0.33\textwidth}
    \centering
    \includegraphics[width=\textwidth]{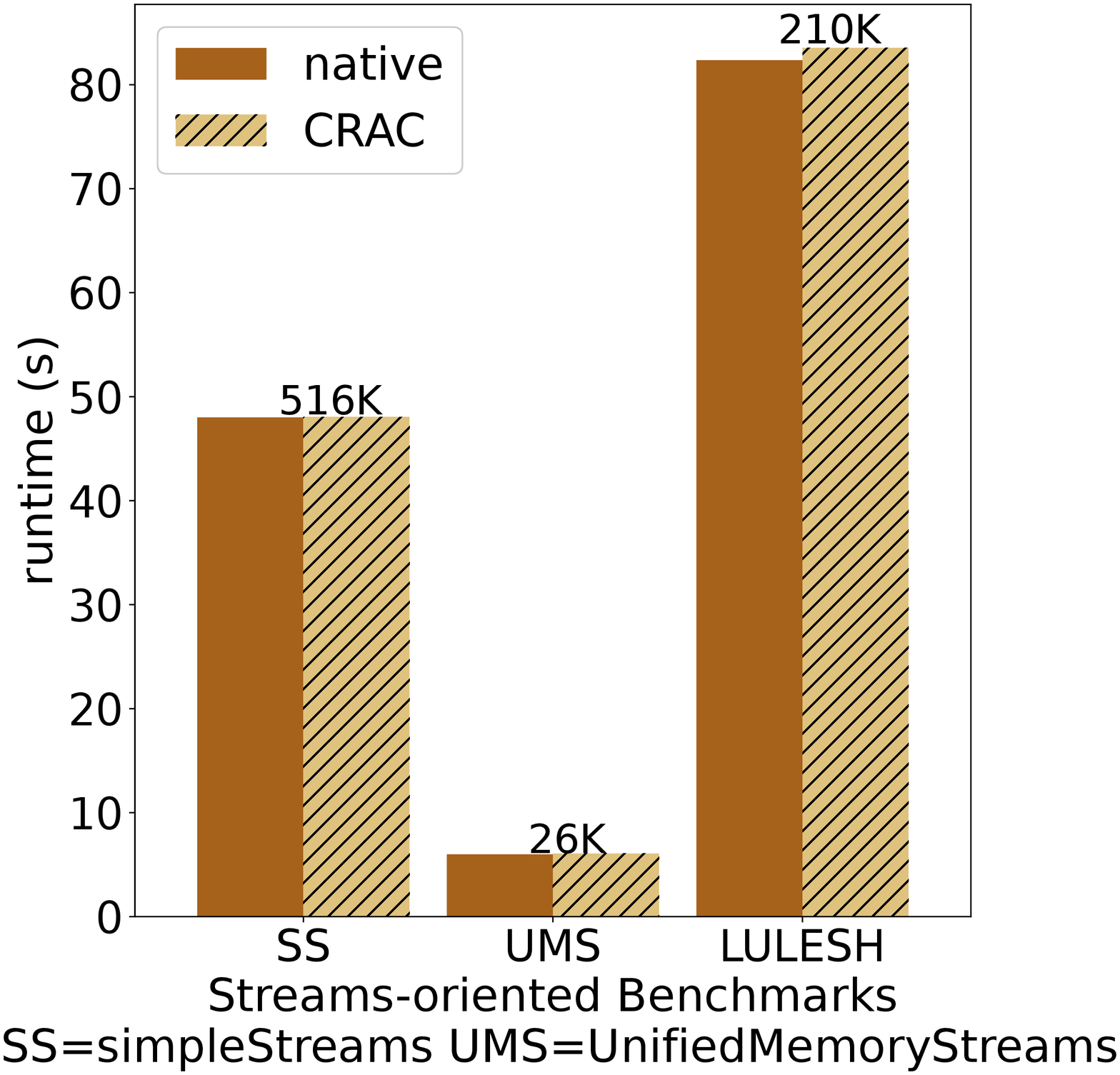}
    \caption{} \label{fig:streams_a}
  \end{subfigure}\hfill
  \begin{subfigure}{0.33\textwidth}
    \centering
    \includegraphics[width=\textwidth]{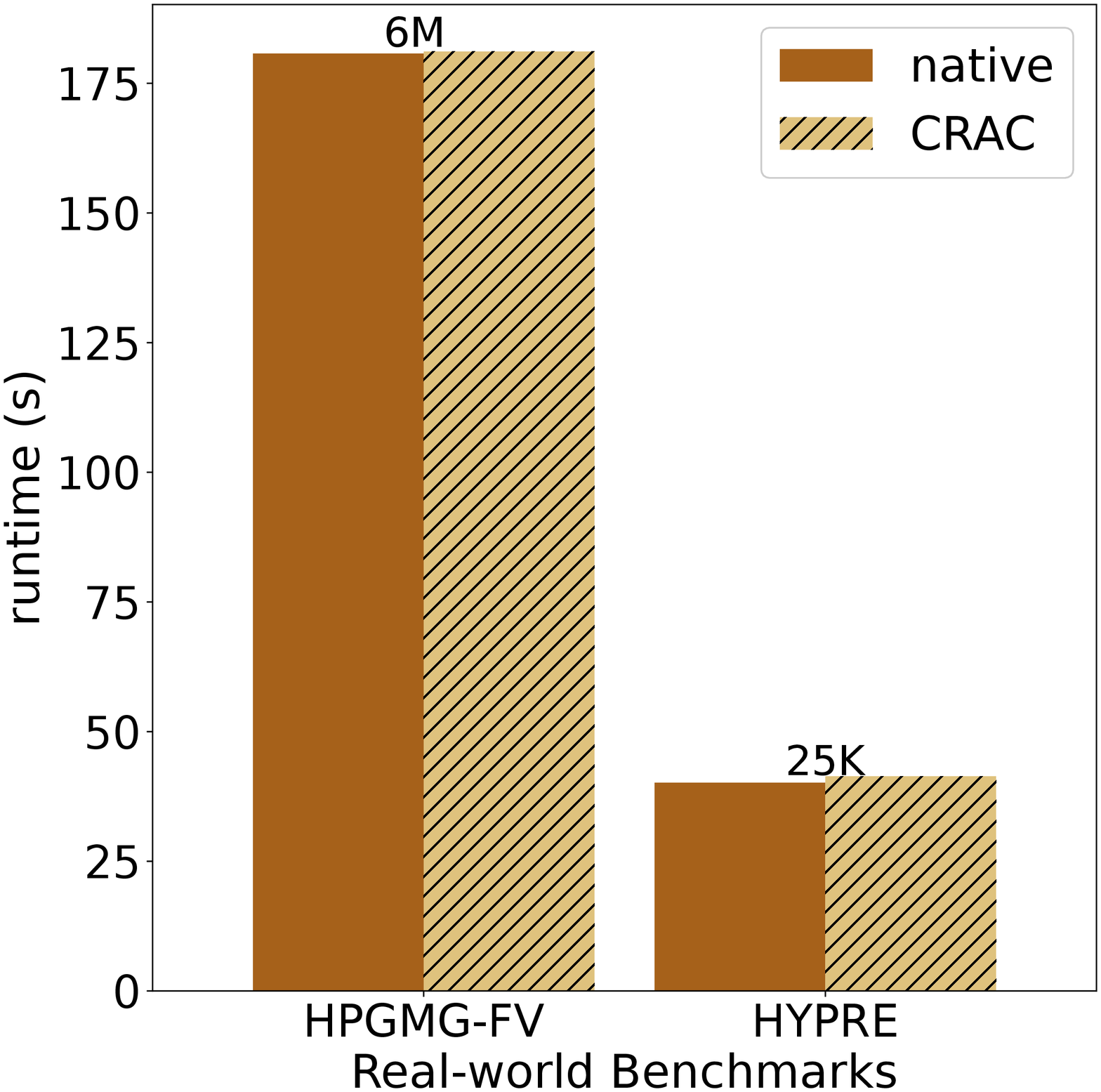}
    \caption{} \label{fig:real_b}
  \end{subfigure}\hfill
  \begin{subfigure}{0.33\textwidth}
    \centering
    \includegraphics[width=\textwidth]{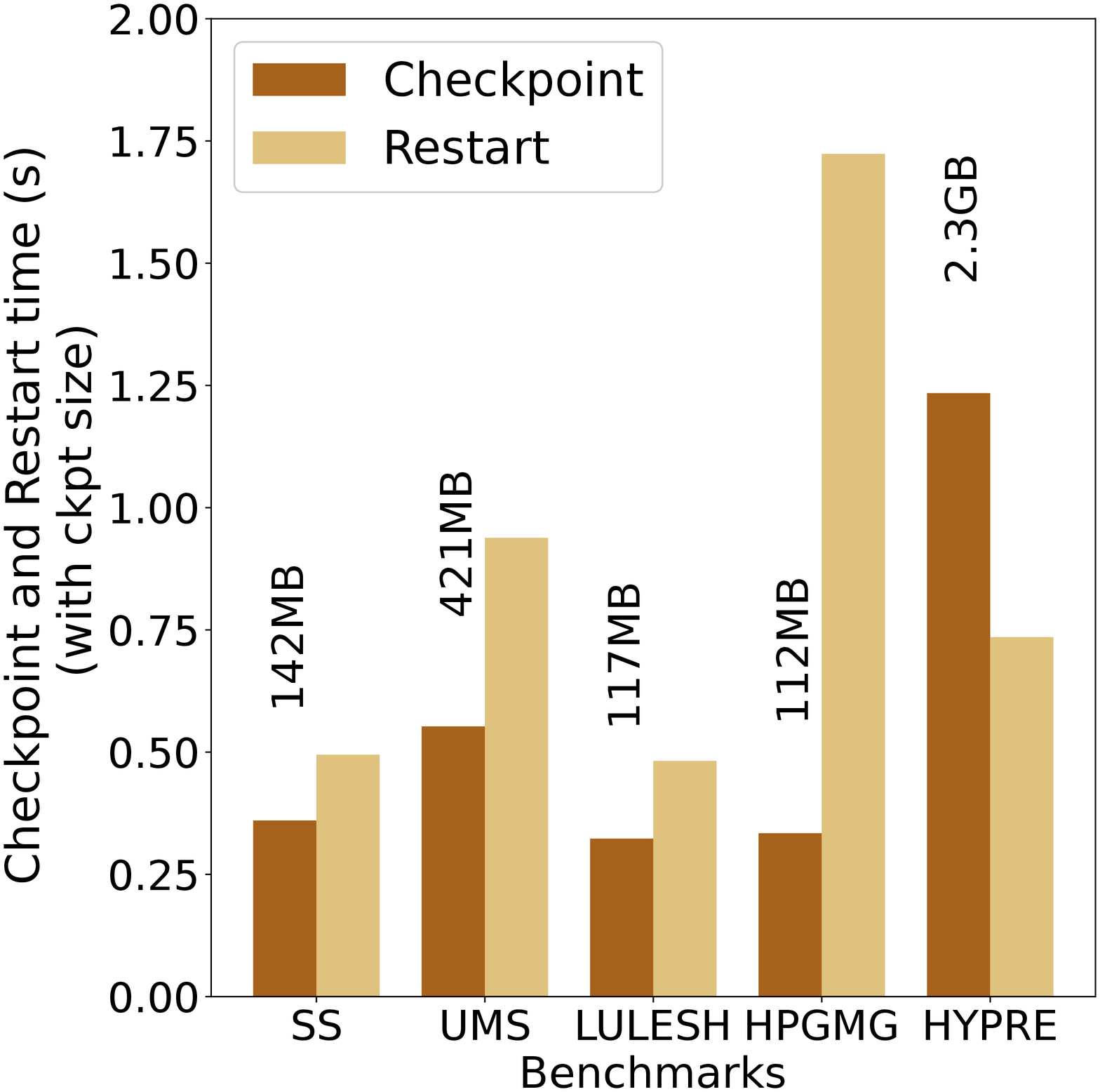}
    \caption{} \label{fig:all_c}
  \end{subfigure}\hfill
  \caption{ Runtimes of (a) Stream-oriented Benchmarks; (b)~Real-world
  Benchmarks (with and without CRAC); and (c)~Checkpoint and Restart times
  with CRAC with checkpoint image size.
  \label{fig:streamsAndRealworldfigs}
  }
\end{figure*}

\paragraph{Configuration and runtime overhead}
  We re-configured the number
  of streams from 4 (default) to 128. For a NVIDIA V100 GPU with its compute
  capability of~7.0, 128 is the maximum concurrent kernel
  limit~\cite{nvidia_max_concurrent_kernel}. The application fails if
  the stream count is increased beyond the max limit. \texttt{nreps} is the
  number of times each experiment is repeated. For better accuracy, we changed
  it from its default value of 10 to 1000. \texttt{niterations} is the
  number of iterations for the loop inside the kernel. We have varied this
  \texttt{niterations} variable with values 5, 10, 100, and 500.  We use
  the default Blocking Sync Event synchronization method. The benchmark
  reports the time to execute one CUDA kernel with streams and without
  streams (i.e., non-streamed).

  Figure~\ref{fig:simpleStreamRuntimes}
  shows how the overall
  runtime of simpleStreams varies with the number of iteration increments.
  CRAC still maintains less than 1\% overhead in each case.
  Figure~\ref{fig:simpleStreamCUDAcalls} (plot on the right) shows the impact
  of CUDA streams.
  As \texttt{niterations} increases (see previous paragraph)
  the time to run the CUDA kernel increases.
  Figure~\ref{fig:simpleStreamCUDAcalls} shows that the streamed version
  becomes significantly faster, compared to the non-streamed version, as
  \texttt{niterations} increases.  Yet CRAC continues to perform with
  low overhead even for the faster streamed version.
  For the same reasons, CUDA
  streams is widely used over regular non-streamed kernel launches. Note that
  CRAC incurs no overhead; neither in non-streamed CUDA kernel execution time
  nor in the one with streams. This shows that even after increasing the
  concurrency level to the max (128 streams), CRAC handles it well as
  compared to previous solutions. Figure~\ref{fig:streams_a} shows the
  runtime with the same configuration (128~streams, 1000~repetitions,
  and 500~iterations).

\paragraph{UnifiedMemoryStreams(UMS)}
UnifiedMemoryStreams (UMS) is taken from NVIDIA's code samples and
illustrates
the usage of streams with Unified Memory. UnifiedMemoryStreams implements a
simple task consumer using threads and streams with all data in Unified Memory,
and tasks consumed by both host and device. The application randomizes task
sizes for a total of 40~tasks with 4~streams. Based on the task size, the
application decides at run time whether the task should be run on the host or
the device. Note that both the device and host are using same unified memory.
\textbf{Configuration and runtime overhead:} We configured the
  application to use 128~streams with a total of 1280~tasks. Since we needed
  to run the experiments 10~times for the average runtime, we set the seed
  to a random number 12701 to get consistent task allocations. We measured
  the execution time by elapsed wall-clock time. Figure~\ref{fig:streams_a}
  shows the average runtime without and with CRAC. We observed an
  overhead of~1.5\%.

\paragraph{LULESH}
Version 2.0 GPU model of LULESH is specifically implemented for NVIDIA's
GPUs.  LULESH is a scientific real-world application developed by Lawrence
Livermore National Laboratory~~\cite{LULESH2:changes} that solves the Shock
Hydronomics Challenge Problem.  LULESH provides two options, one with
an unstructured grid and the other with a structured grid. In our case,
we use a structured grid with 150 edge elements, which makes the
problem size $150\times150\times150$, and which uses nearly 2~GB of memory.

\textbf{Runtime overhead:} LULESH calls 210K~CUDA calls in 80~seconds
  of its execution time that means around 2.5K CUDA calls per second. We
  saw that with maximum streams in simpleStreams and UnifiedMemoryStreams,
  CRAC still incurs low-overhead. With the real-world application that
  makes 65K~cudaLaunchKernel calls. Figure~\ref{fig:streams_a} shows
  that LULESH's performance is still the same with CRAC, with an overhead
  slightly less than 2\%.

\begin{figure*}[ht!]
  \centering
  \begin{minipage}{0.45\textwidth}
    \centering
    \includegraphics[width=\textwidth]{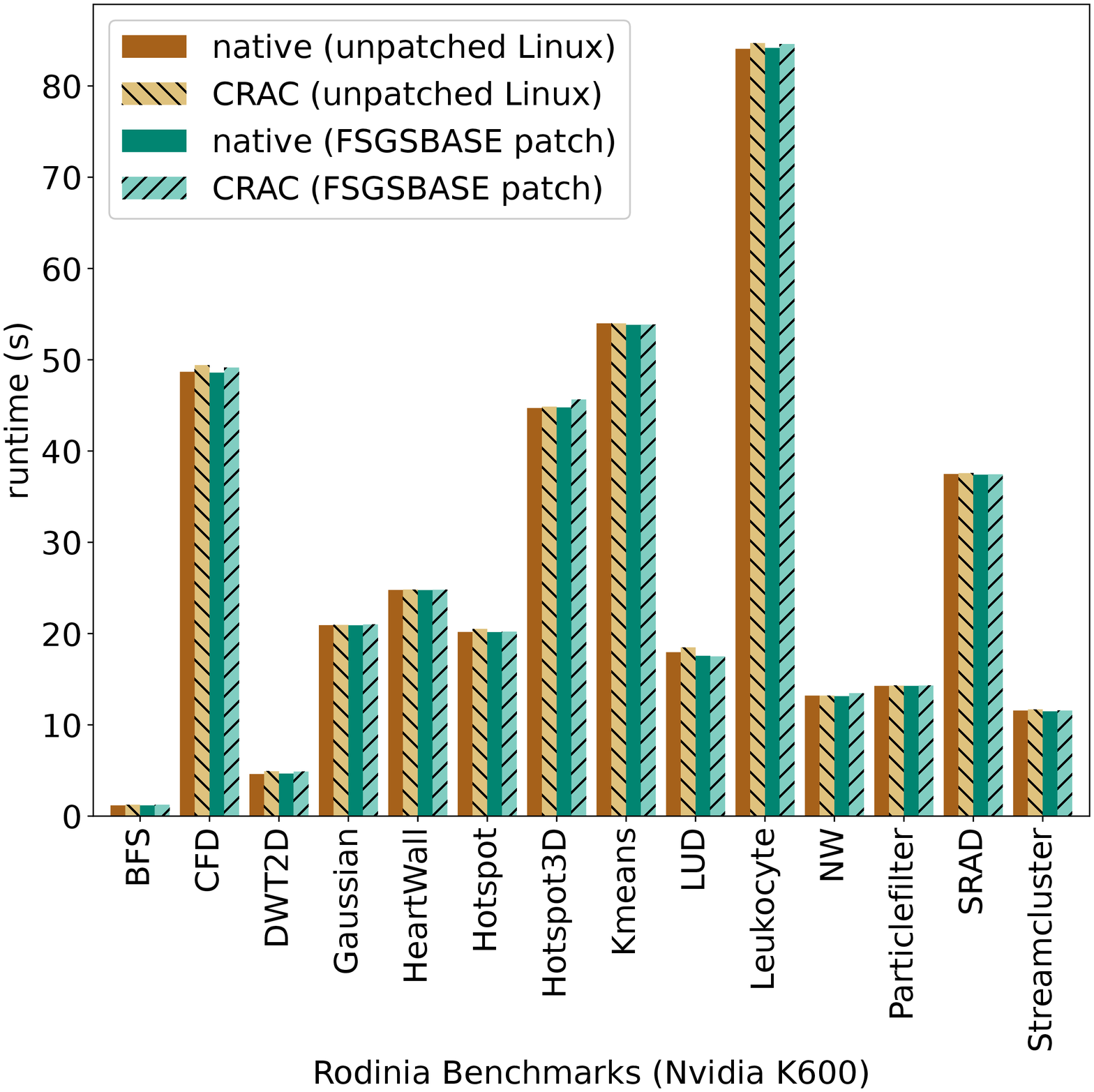}
  \end{minipage}
  \hspace{0.3in}
  \begin{minipage}{0.45\textwidth}
    \centering
    \includegraphics[width=\textwidth]{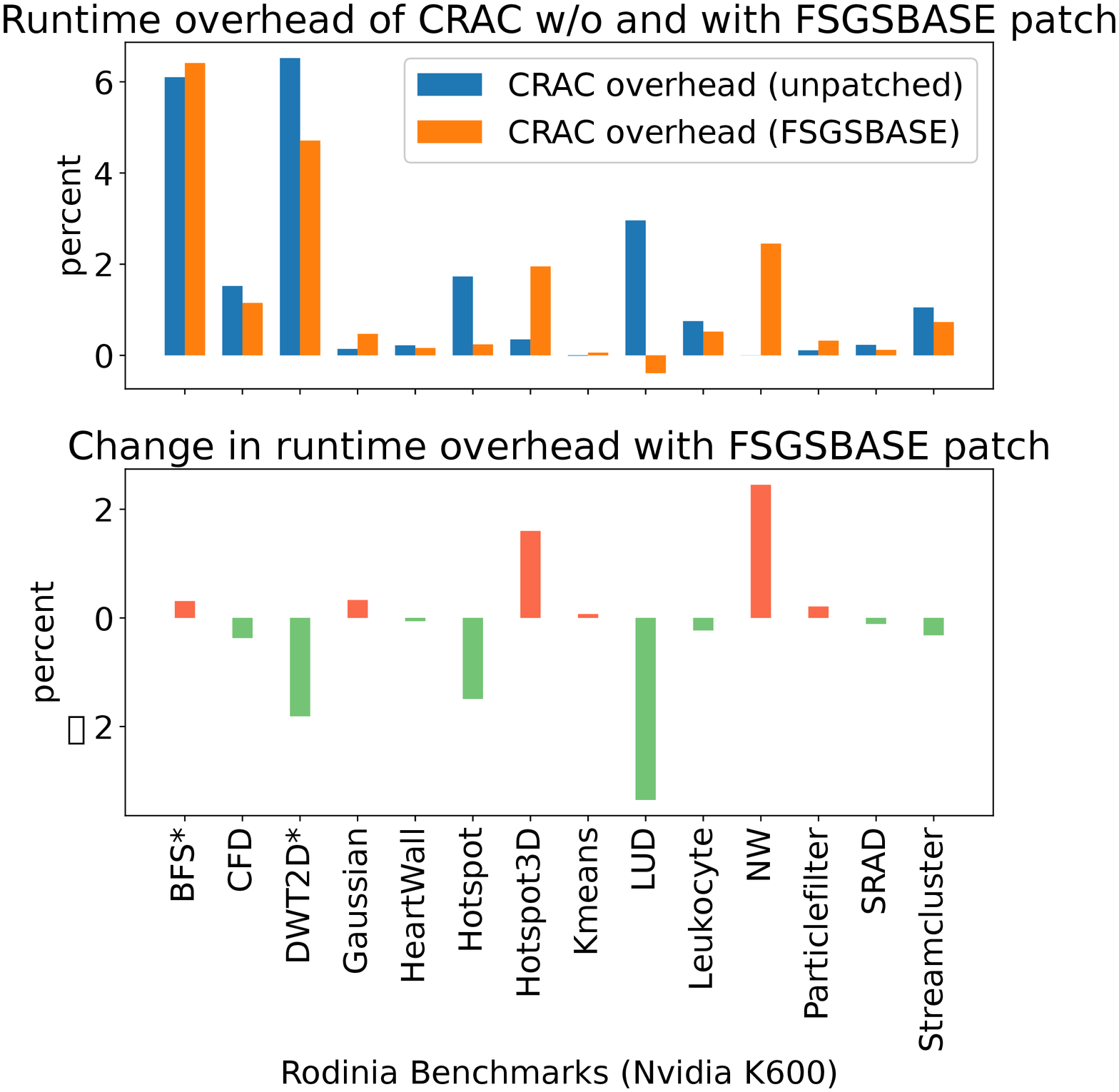}
  \end{minipage}
\caption{(left) Runtimes of Rodinia Benchmarks without and with CRAC on
         both unpatched and \\
         \hbox{~~~~~~~~~~~~~~~~~~~~} patched (FSGSBASE) Linux; \\
         \hbox{~~~~~~~~~~~~}(right, top) Runtime
         overhead of CRAC on both unpatched and \\
         \hbox{~~~~~~~~~~~~~~~~~~~~} patched (FSGSBASE) Linux; \\
         \hbox{~~~~~~~~~~~~}(right, bottom) and percentage difference observed
         with CRAC's runtime overhead with \\
         \hbox{~~~~~~~~~~~~~~~~~~~~} patched Linux as compared to
         unpatched Linux (lower is better).
\label{fig:fsgs}}
\end{figure*}

\paragraph{Checkpoint overhead} Figure~\ref{fig:all_c} shows that the
checkpoint overhead is very low as compared to the overall runtime of
each stream-oriented application.
CRAC needs to recreate streams and make the CUDA library's state consistent. So, the time
is slightly more than the checkpoint time.  However, both checkpoint and restart finish
within one second in each stream-oriented application.

\subsubsection{\textbf{Real-world applications (HPGMG-FV and HYPRE)}}

HPGMG is a high-performance geometric multigrid application.  It is
one of the benchmarks used for ranking speeds of the top
supercomputers~\cite{hpgmg_benchmarking}. We use HPGMG-FV (Finite Volume) for
our experiments. HPGMG-FV can be scaled further with MPI over multiple nodes.
However, it suffices for our purposes to run HPGMG-FV over a single MPI rank.
This already provides a real-world scale since this configuration of HPGMG-FV
make 2~million CUDA calls
per minute (35,000~CUDA calls per second).  This scale is already representative
of real-world high-performance applications.

HYPRE is a linear system solver library that makes only 600~CUDA calls per
second. However, HYPRE creates large UVM regions, and employs long-running kernels.
One MPI-rank can create UVM regions of up to 1~GB, and the host and
the device both work simultaneously on UVM regions via CUDA streams and textures.
Therefore, HYPRE incurs a higher memory footprint than HPGMG-FV. The following
table shows the command-line arguments needed to run these two real-world
applications.

\begin{table}[ht]
  \centering
  \begin{tabular}{ |l|l| }
    \hline
    \textbf{Application} & \textbf{Command-line arguments}\\ \hline
    \texttt{HPGMG-FV} & 7 8 \\ \hline
    \texttt{HYPRE} & ij -solver 1 -rlx 18 -ns 2 -CF 0 -hmis -interptype 6 \\
                   & -Pmx 4 -keepT 1 -tol 1.e-8 -agg\_nl 1 \\
                   & -n 250 250 250 250 \\
    \hline
\end{tabular}
\end{table}

\textbf{Runtime overhead:} Figure~\ref{fig:real_b} shows native and
  CRAC runtimes for both HPGMG-FV and HYPRE.  Figure~\ref{fig:real_b} shows
  an average of 10~native and 10~runs with CRAC.
  CRAC manages to run both the application with barely
  less than 2\% with HPGMG and 3\% with HYPRE.

\textbf{Checkpoint overhead:}
  Figure~\ref{fig:all_c} shows that with HPGMG, CRAC need to replay a
  lot of CUDA APIs as compared to its memory footprint.  Therefore, CRAC
  takes nearly 1.75~seconds to restart HPGMG.  On the other hand, HYPRE's
  checkpoint size is of 2.3GB but it takes less time to restart.  With all
  the results we have seen so far, one can conclude that the runtime
  overhead with CRAC is very low and checkpoint overhead is not much.
  However, the restart time can be larger, depending on how many CUDA
  calls CRAC need to replay to make newer CUDA library's state consistent.

\subsubsection{\textbf{Comparison of CRAC to Proxy-based
  approaches:  The cost of IPC}}

As described at the beginning of Section~\ref{sec:design}, the starting
point in the new approach of CRAC was the observation that the existing
proxy-based approaches to checkpointing CUDA (e.g., by CRCUDA and CRUM)
rely on expensive inter-process communication (IPC) between CUDA application
and the proxy.  In real-world experiments, the authors of CRUM measured the
runtime overhead on real-world benchmarks at from 6\% to 12\%.

Here we present a synthetic IPC benchmark
(CMA/IPC in Table~\ref{tab:ipc-experiments}) and compare with native CUDA and
CRAC.  CMA is Cross-Memory Attach (i.e., the Linux syscalls
process\_vm\_readv and process\_vm\_readv).
It is based on BLAS~\cite{blackford2002updated}
(Basic Linear Algebra Subprograms), and uses the NVIDIA cuBLAS library.
The cuBLAS library resides in the lower half and is directly called
from the upper half.  In the case of CMA/IPC, the buffers are copied
via CMA from the application to a proxy process (which executes the
cuBLAS routine), and the result is copied back to the application.

We ran three types of programs: cublasSdot (inner product),
cublasSgemv (matrix-vector product), and cublasSgemm (matrix-matrix product).
The dimension was chosen so that the matrix (or vector, for cublasSdot)
had data size 1~MB, 10~MB, or 100~MB.  The programs called the
respective cuBLAS routines 10,000 times, as part of a timing loop.
Times in milliseconds are reported for a single iteration.

\begin{table}[ht]
	\centering
	\begin{tabular}{ |l|r|r|r|l| }
		\hline
		\textbf{CUDA Call} & Data &
                \textbf{Native} &
                \textbf{CRAC(ms)} &
		\textbf{CMA/IPC(ms)}\\
		\textbf{} & size & \textbf{(ms)} &
                \textbf{(\% overhead)} &
		\textbf{(\% overhead)}\\
\hline

\texttt{cublasSdot} & 1MB   & 0.026  & 0.027~(3.9)   & ~~0.21~(698) \\
\hline
\texttt{cublasSdot} & 10MB  & 0.049  & 0.050~(3.3)   & ~~2.56~(5142) \\
\hline
\texttt{cublasSdot} & 100MB & 0.282  & 0.284~(0.5)   & ~50.4~~(17766) \\
\hline
\texttt{cublasSgemv} & 1MB  & 0.012  & 0.012~(1.9)   & ~~0.082(577) \\
\hline
\texttt{cublasSgemv} & 10MB & 0.036  & 0.037~(0.7)   & ~~1.25~(3329) \\
\hline
\texttt{cublasSgemv} & 100MB& 0.142  & 0.142(-0.1)  & ~25.5~~(17812) \\
\hline
\texttt{cublasSgemm} & 1MB  & 0.202  & 0.207~(2.4)   & ~~0.49~(142) \\
\hline
\texttt{cublasSgemm} & 10MB & 1.806  & 1.816~(0.6)   & ~~9.03~(400) \\
\hline
\texttt{cublasSgemm} & 100MB& 32.373 & 32.107~(-0.8) & 100.34~(209) \\
\hline
   \end{tabular}
	\caption{Comparison of CRAC to an IPC-based approach
                 (e.g., as in CRCUDA and CRUM)
	\label{tab:ipc-experiments}
        }
\end{table}

Note that the CRAC overhead has generally about 1\% overhead.  The CRAC
overhead can rise to 3.9\% for a dot product of vectors
of 1~MB size.  We attribute this to cache effects.

In comparison, the overhead using CMA~\cite{vienne2014benefits} (Cross
Memory Attach) for IPC varies from 142\% to 17,812\%.  The overhead
is huge, as expected.  To be fair, CRCUDA and CRUM were intended to
run on real-world programs in which buffers could be less than 1~MB
and where the computation is not dominated by very frequent CUDA calls.
(For cublasSdot, there are 1/(0.026~ms) = 38,000 calls per second.)

\subsubsection{\textbf{Runtime overhead improvement using Linux's
  upcoming FSGSBASE patch}}
\label{sec:fsgsbase}

A small experiment was also performed to see if there was significant
benefit to using the upcoming FSGSBASE
patch to the Linux kernel~\cite{fsgsbase-v9}.  In the current Linux,
switching to a new thread (or to the lower-half program in our case)
requires a kernel call to set the corresponding x86-64 ``fs'' register
for that thread.  A kernel call may require a millisecond.  If done
frequently, this can be expensive.  At least in the case of MPI applications,
it was previously observed in~\cite{garg2019mana} that the expense
of the kernel calls was significant when calling lower-half routines.

Hence, we wished to see if CRAC's already small runtime overhead could be
further reduced by using the FSGSBASE patch to directly set the ``fs''
register, instead of setting ``fs'' through kernel calls.  As we shall
see below, the added advantage of using the FSGSBASE patch is small,
and often nearly zero.

To test this question, we
analyze whether CRAC's runtime overhead can be further reduced using the FSGSBASE patch.
CRAC needs to get and set the ``fs'' register when it makes a call from
the upper half to the lower half.  (This is analogous to the use of the
``fs'' register in context switches among threads in Linux.)  Setting the
``fs'' register is expensive due to the kernel call.

The experiments of Figure~\ref{fig:fsgs} were run on a local node: an
older NVIDIA Quadro K600 GPU.  It was not possible to install a patched
Linux kernel on the public, production nodes used for the experiments in
the other figures.
This also explains why the same Rodinia benchmarks mostly ran for at
least 10~seconds in this experiment.

Figure~\ref{fig:fsgs} presents two columns of graphs. On the left, the original
14~Rodinia benchmarks are plotted.  Each benchmark shows the native
runtime and the CRAC runtimes, both with and without the FSGSBASE patch.
The runtimes with FSGSBASE were taken using the FSGSBASE/v9 kernel
patches~\cite{fsgsbase-v9}.

The two graphs on the right in Figure~\ref{fig:fsgs} present the same
data, but they express the data as percentage differences, to more
clearly contrast two cases:  the runtime overhead of CRAC (with and
without FSGSBASE); and the change in runtime overhead of CRAC when using
the FSGSBASE patch.  Lower is better in both cases.

\section{Related Work}
\label{sec:relatedWork}

Much of the work targeting transparent checkpointing of
CUDA was already covered in Section~\ref{sec:background}, as part of
the motivation for a fresh approach in CRAC.
See Section~\ref{sec:background} for more details.

To summarize, several
techniques~\cite{shi2009vcuda,gupta2009gvim,takizawa2009checuda,
gomez2010transparent,nukada2011nvcr} were explored prior to CUDA~4.0 (in
2011 and earlier).  Later, unified memory between device and host was
introduced to CUDA in two increments: Unified Virtual Addressing (UVA)
in CUDA~4.0; and Unified Virtual Memory (UVM~\cite{sakharnykh_gtc_2017})
in CUDA~6.0.  This unified memory was incompatible with existing
checkpointing approaches.

Since then, two newer checkpointing approaches appeared:
CRCUDA~\cite{suzuki2016transparent} and CRUM~\cite{garg2019mana}.
The limitations of these two approaches were
already described:  high runtime overhead,
incomplete UVM support, and untested scaling of concurrent streams.
(See the second page of Section~\ref{sec:introduction} for details.)

It remains to describe four techniques from the literature that
are related to the implementation of CRAC:  proxies in CRUM;
proxies in the wider literature; split processes; and
process-in-process.

\paragraph{Use of proxy processes in CRUM}

The previous work of CRUM in checkpointing CUDA has unacceptable high
overhead of 6\% and could go up to 12\%. This occurs at two extremes.

\begin{description}
\item[Case I:]{
  ~~~{\em Many short-lived kernels.}  This incurred overhead because
  of the need to frequently marshal and unmarshal the parameters for
  communication between the application and the proxy process when
  invoking CudaLaunchKernel. For example, HPGMG-FV has a high frequency
  of CUDA calls.
}
\item[Case II:]{
  ~~{\em Kernel and host access many UVM memory pages frequently.}
  This requiring frequent calls to \texttt{mprotect} and userfault\_fd
  (a recent Linux utility serving the same purpose as segfault
  handlers). This interacted particularly badly with NVIDIA UVM.
}
 \end{description}

\paragraph{Proxy processes}
{\em Proxy processes} are a well-known concept that is widely used in systems.
In an early example, Zandy et~al.~\cite{zandy1999process}
demonstrated the use of a ``shadow'' process for checkpointing currently
running application processes that were not originally linked with a
checkpointing library. This allows the application process to continue to
access its kernel resources, such as open files, via RPC calls with the
shadow process.
Kharbutli et~al.~\cite{kharbutli2006comprehensively} use a proxy process
for isolation of heap accesses by a process and for containment of attacks
to the heap.
CheCL~\cite{takizawa2011checl} has employed
proxy processes already in 2010, for the closely related OpenCL
language~\cite{stone2010opencl} for GPUs.
CRCUDA~\cite{suzuki2016transparent} and CRUM also
employ proxy processes.

\paragraph{Split processes}

Split processes were described in
Figure~\ref{fig:split-process} in Section~\ref{sec:split-processes}.
MPI for MANA~\cite{garg2019mana} had adopted the idea of split
processes in the context of checkpoint-restart for MPI.  Upon checkpoint, only
the upper half memory is saved.  On restart, a small bootstrap program
in the lower half restores the upper half memory, and the upper half
then replays any persistent state associated with a physical device.
In the case of MANA, that physical device would be the network, and/or
sockets communicating with a central MPI coordinator.  In the case of
the current work (CRAC), the physical device is the GPU.

There are several antecedents to the idea of combining two programs
in a single process.  Here we note McKernel and shadow device drivers,
both devised for the Linux kernel.

McKernel~\cite{gerofi2016scalability} runs a ``lightweight'' kernel along
with a full-fledged Linux kernel. The HPC application runs on the
lightweight kernel, which implements time-critical
system calls. The rest of the functionality is offloaded to a proxy process
running on the Linux kernel. The proxy process is mapped in the address
space of the main application, similar to MANA's concept of
a lower half, to minimize the overhead of ``call forwarding'' (argument
marshalling/un-marshalling).

Swift \hbox{et~al.}~\cite{swift2006recovering} developed the idea of
a ``shadow device driver''.  The lower half corresponds to the actual
device driver, and the upper half corresponds to a shadow device
driver that mirrors (or ``logs'') all transactions to the lower half.
If the lower-half device driver crashes, then it is re-initialized
and a long-and-replay approach is used to re-initialize it.

\paragraph{Process-in-process:  an approach related to split processes}

Process-in-process~\cite{hori2018process} is related to split
process in that
that both approaches load multiple programs into a single address
space.  However, the goal of process-in-process was intra-node communication
optimization, and not checkpoint-restart.
Given two MPI ranks (processes) co-located on a single computer node,
the two ranks were loaded into a single address space, to make
copying of messages between the two MPI ranks more efficient.

Unlike split processes, process-in-process
loads {\em all} MPI ranks co-located on the same node
as separate {\em threads} within a single process, but
in different logical ``namespaces'', in the sense of the \texttt{dlmopen}
namespaces in Linux.

\section{Conclusion and Future Work}
\label{sect:conclusion}

Transparent checkpointing of CUDA with low runtime overhead has been
demonstrated.  This is important, since most earlier checkpointing
approaches (prior to CUDA~4.0) are incompatible with the versions of CUDA
introduced in version~4.0 and beyond.  There do exist two other recent
solutions that are compatible with CUDA versions~4.0 and later.  But both
have limited functionality, and more importantly, their runtime overhead
is very high (e.g., 6\% to 12\% on complex, real-world applications).
This is an unacceptable waste of resources for the expensive GPUs used
in mid-level computing and supercomputing.

The current solution demonstrates low runtime overhead (about~1\%), and
additionally, that low overhead is maintained for two
important, advanced CUDA features.  The two features are:
(a)~support for the maximum capability of parallel GPU streams;
and (b)~full support for CUDA's UVM.
The parallel GPU-stream support includes near-native runtime performance.
CUDA streams are gaining importance, for example, in multi-threaded
programs on many-core CPUs, in which each thread employs a separate
CUDA stream.  The Full support for UVM is important in simplifying
software development in CUDA.

Some extensions of this work have also been demonstrated, which have
the side benefit of providing a roadmap toward future extensions.
For example, the real-world applications used in this paper not only use
the CUDA runtime library, but also other CUDA libraries such as cuBLAS and
cuSolver.  Hence, the current work can easily be extended to support other
CUDA libraries and additional GPU APIs.  Further, a proof of principle
was demonstrated for checkpointing of hybrid MPI+CUDA on a single node.
In future work, this proof of principle for transparent checkpointing
of MPI+CUDA will be extended to full support for MPI on multiple nodes.

\section*{Acknowledgment}
We thank Michael Sullivan of NVIDIA for a careful reading and comments,
and also for the use of computer resources at NVIDIA.  We also thank Rohan
Garg for conversations describing his earlier design of CRUM for CUDA.

\bibliographystyle{alpha}
\bibliography{dmtcp-cuda-split}

\end{document}